\documentclass[twoside,9pt]{article}
\usepackage{extsizes}
\usepackage[super,sort&compress,comma]{natbib} 
\usepackage[version=3]{mhchem}
\usepackage[left=1.5cm, right=1.5cm, top=1.785cm, bottom=2.0cm]{geometry}
\usepackage{balance}
\usepackage{mathptmx}
\usepackage{tcolorbox}
\usepackage{etoolbox}
\usepackage{sectsty}
\usepackage{graphicx} 
\usepackage{lastpage}
\usepackage[format=plain,justification=justified,singlelinecheck=false,font={stretch=1.125,small,sf},labelfont=bf,labelsep=space]{caption}
\usepackage{float}
\usepackage{fancyhdr}
\usepackage{fnpos}
\usepackage[english]{babel}
\addto{\captionsenglish}{%
}
\usepackage{array}
\usepackage{droidsans}
\usepackage{charter}
\usepackage[T1]{fontenc}
\usepackage[usenames,dvipsnames]{xcolor}
\usepackage{setspace}
\usepackage[compact]{titlesec}
\usepackage{hyperref}
\usepackage{subcaption}
\usepackage{siunitx}

\usepackage{epstopdf}%This line makes .eps figures into .pdf - please comment out if not required.
\definecolor{cream}{RGB}{222,217,201}
\titleformat{\section}          % Format for \section
  {\bfseries\fontsize{12pt}{14pt}\selectfont} % Bold, 12pt font
  {\thesection}                % Section number
  {1em}                        % Space between number and title
  {}                           % Before the title
\begin{document}

%%%TITLE AND AUTHORS%%%
%\title{Title} %The title of the article should be inputed there
%\author{Author Full Name,*a Author Full Name b and Author Full Name c}
%\date{} %do not delete or the date will appear in the generated pdf
%\maketitle
\begin{flushleft}
{\fontsize{18pt}{20pt}\selectfont\textbf{Speciation and hydration forces in sodium carbonate/bicarbonate aqueous solutions nanoconfined between mica sheets$^\dag$}}
\end{flushleft}
{\fontsize{12pt}{14pt}\selectfont Daria Turculet, Shurui Miao, Kieran J. Agg and Susan Perkin\textsuperscript{*}}\\\\
Received 00th January 20xx, Accepted 00th January 20xx
DOI: 10.1039/x0xx00000x\vspace{0.5cm}
%%%END OF TITLE AND AUTHORS%%%
\section*{Abstract}
The equilibrium between hydrated and hydrolysed forms of \ce{CO2} in water is central to a multitude of processes in geology, oceanography and biology. Chemistry of the carbonate system is well understood in bulk solution, however processes such as mineral weathering and biomineralisation frequently occur in nano-confined spaces where carbonate chemistry is less explored. For confined systems, the speciation equilibria are expected to tilt due to surface reactivity, electric fields and reduced configurational entropy. In this discussion paper we provide measurements of interaction force between negatively charged aluminosilicate (mica) sheets across aqueous carbonate/bicarbonate solutions confined to nanoscale films in equilibrium with a reservoir of the solution. By fitting the measurements to a Poisson-Boltzmann equation modified to account for charge regulation at the bounding walls, we discuss features of the bicarbonate speciation in confinement.  We find that (i) the presence of bicarbonate in the bulk reservoir causes a repulsive excess pressure in the slit compared to pH-neutral salt solutions at the same concentration, arising from a higher (negative) effective charge on the mica surfaces; (ii) the electrostatic screening length is lower for solutions of \ce{Na2CO3} compared to \ce{NaHCO3} at the same bulk concentration, due to a shift in the speciation equilibria with pH and in accordance with Debye-H\"{u}ckel theory;  (iii) hydration forces are observed at distances below 2~nm with features of size 0.1~nm and 0.3~nm; this was reproducible across the various bicarbonate electrolytes studied, and contrasts with hydration forces of uniform step size measured in pH-neutral electrolytes. 

\vspace{1cm}
\section*{Introduction}
Carbonates play a key role in regulating Earth’s carbon cycle and often serves as reliable record of the co-evolution of Earth and life. The geological formation of carbonate minerals is highly sensitive to variations in calcium and magnesium concentrations. Even subtle changes in the relative abundance of these ions can shift the delicate balance of electrostatic interactions, hydration forces, and mineral formation and dissolution kinetics, favouring the precipitation of specific polymorphs such as calcite, aragonite, or dolomite.\cite{Coevolution_WilliamsRickaby2012,Dolomite_Sun2023} Likewise, life on Earth has evolved intricate and diverse mechanisms to control carbonate mineralization. Organisms such as corals, mollusks, and plankton actively mediate carbonate mineral formation by regulating local ion concentrations and pH, enabling the construction of shells and skeletons.\cite{Biomin_LowenstamWeiner1989} In addition to the structural role of biomineralisation, the carbonate-bicarbonate system is also one of the most fundamental acid-base regulation systems in biology. It is widespread among life forms ranging from vertebrates and microbes, where it serves as a buffer system to maintain pH balance\cite{Biochem_VoetPratt2016}; to algae and plants, where it is used to produce dissolved \ce{CO2} as an alternative carbon source for photosynthesis\cite{AlgaeCO2_Raven2005}. \\

In aqueous environments, sodium carbonate and bicarbonate solutions constitute an acid-base buffer system. Dissolved inorganic carbon (DIC) does not exist as a single species but as a dynamic equilibrium among carbonic acid (\ce{H2CO3}), bicarbonate (\ce{HCO_3^-}), and carbonate (\ce{CO3^{2-}}), whose relative abundances are governed by the bulk pH through the following equilibria (at $T = 298~\mathrm{K}$):

\begin{align}\label{eqn:bicarbonate}
\mathrm{H}_2\mathrm{CO}_3 \mathrm{(aq)}+ \mathrm{H}_2\mathrm{O} \mathrm{(l)}
&\rightleftharpoons
\mathrm{H}\mathrm{CO}_3^- \mathrm{(aq)}+ \mathrm{H}_3\mathrm{O}^+\mathrm{(aq)} &{K_{1}} = 4.3 \times 10^{-7} \nonumber\\
\mathrm{H}\mathrm{CO}_3^- \mathrm{(aq)}+ \mathrm{H}_2\mathrm{O} \mathrm{(l)}&\rightleftharpoons \mathrm{CO}_3^{2-} \mathrm{(aq)}+ \mathrm{H}_3\mathrm{O}^+\mathrm{(aq)} &{K_{2}} = 4.8 \times 10^{-11}
\end{align}

\begin{figure}[h!]
    \centering
    \includegraphics{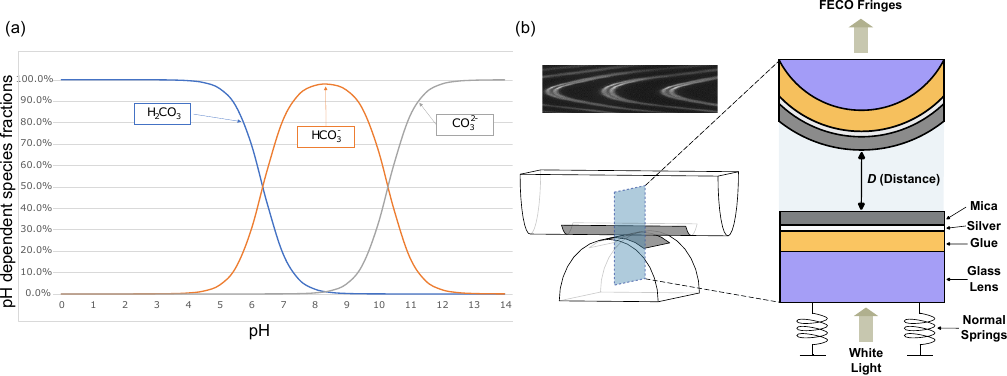}
    \hfill
    \caption{(a) Fractional concentrations of the three dissolved inorganic carbon (DIC) species present in bicarbonate-containing solutions as a function of electrolyte pH. Calculated using the equilibrium constants as in equation~\ref{eqn:bicarbonate} as shown in the SI. (b) Schematic diagram of the SFB used to measure interactions between mica sheets across electrolyte solution. The mica-coated lens setup is immersed in bulk solution, so that confined system is in equilibrium with a reservoir.}
    \label{fig:1}
\end{figure}

The equilibrium constants\cite{crc_handbook_acidity_constants} (or acid dissociation constants) determine the speciation of carbon: carbonic acid dominates under acidic conditions, bicarbonate prevails in the near-neutral range, and carbonate ions become significant at high pH. The fractional abundances of each species as a function of pH are illustrated in Figure \ref{fig:1}(a), which highlights the transition from \ce{H2CO3} to \ce{HCO_3^-} and finally to \ce{CO3^{2-}} with increasing pH. 

~\\
Many geological and biological processes involving carbonates, such as weathering and biomineralisation, occur within confined environments. The interfaces and confining geometry imposed by mineral pore spaces or biomolecular matrices such liposomes or extracellular fluid compartments direct reactivity towards distinct pathways\cite{Biomin_LowenstamWeiner1989}. Such confinement appears to strongly influence the nucleation, growth, and dissolution of carbonate\cite{Crystgrow_MeldrumChristenson2010,CrystGrow_Dysthe2022,CrystGrow_Dziadkowiec2024}, however the mechanisms by which this occurs are complex and not fully known\cite{CaCO3Speciation_Mundy2024,CrystGrow_Meldrum2025}. Electric fields arising from surface charges, specific chemical interactions with the confining surfaces, and restricted configurations and limited ion mobility in confined regions are all likely to play a role. Understanding how charged solid-liquid interfaces, and degrees of confinement, shape the thermodynamic and kinetic behaviour of carbonate-bicarbonate systems is therefore essential for linking molecular processes to macroscopic geochemical and biological outcomes.\\

In this work we present model experiments representing an idealised confined geometry: we study aqueous solutions of the carbonate/bicarbonate system confined to nanoscale films between two atomically smooth muscovite mica sheets using a surface force balance (SFB)\cite{Hayler_2024}. Muscovite mica is well suited to high-resolution experimental studies of this sort due to its ideal cleavage along the basal plane revealing an atomically-smooth crystal surface over macroscopic ($\approx$ cm$^2$) areas.  Muscovite mica is a phyllosilicate clay with overall composition \ce{KSi3Al3O10(OH)2} consisting of tetrahedral-octahedral-tetrahedral (TOT) silicate layers. 1 out of 4 of the Si atoms in the tetrahedral layers is substituted for Al, leading to a net negative charge on the TOT lattice which is neutralised by \ce{K+} ions lying between the aluminosilicate sheets\cite{Christenson2016}. When immersed in electrolyte solutions, the surface \ce{K+} ions can exchange with cations from the bulk electrolyte. The adsorbed cations are either directly associated with the aluminosilicate lattice or retain their primary hydration layer; the latter form the so-called Stern layer\cite{Fenter2012}. However, ions adsorbed (hydrated or otherwise) to mica in aqueous solution usually do not fully neutralise the surface charge; instead, the surface retains and effective charge which is neutralised by dissolved ions in nearby solution extending over a distance characterised by the electrostatic screening length ($\kappa_\textrm{D}^{-1}$)\cite{Smith2020}. According to mean-field descriptions, two particles or surfaces approaching to distances such that the electrical double layers overlap will experience an interaction force, always repulsive for two identical surfaces, arising from the excess osmotic pressure in the confined region.  Measurements of this force/pressure can be fitted (in the far-field) using the Poisson-Boltzmann equation with appropriate boundary conditions; the result is fitted empirical values for the effective surface potential or surface charge, screening length, dielectric permittivity, etc. under different solution conditions. \\
 
At very strong confinement (to surface separations below a few nm), an additional \textit{hydration force} is often measured between two mica sheets across electrolyte solutions\cite{Israelachvili1983}. In previous measurements, it was found that the nature of the hydration forces is strongly connected to the ion exchange process. For example, in acidic solutions there were no hydration forces at all; face-to-face mica sheets experience an attraction into direct contact as predicted by DLVO theory. On the other hand, in solutions containing alkali halide salts (LiCl, NaCl, KCl, CsCl) above a critical concentration an additional short-range repulsion was measured. These additional short-range forces, or hydration forces, were attributed to the adsorption of hydrated ions at the mica surface which are not de-hydrated under the attractive van der Waals pressure between the mica sheets\cite{Pashley1981a,Pashley1981b,Mugele2019}. Detailed investigation of the hydration forces revealed that the repulsion is not monotonic, but instead consists of oscillations with a wavelength close to the size of a water molecule\cite{Israelachvili1983,Hallett2023,Agg2025}. The observation of oscillatory hydration force has been accompanied by a heuristic interpretation of ‘squeezing out’ water layers, while theoretical models more rigorously indicate that oscillations arise from a resonance of the bulk structure in the cavity between the surfaces\cite{Hedley2023,Groves2024}. \\

Measurements of interaction forces between mica sheets across a wide range of electrolyte solutions have been made in the past\cite{Pashley1981a,Pashley1981b,Valtiner2014}, revealing trends in the surface adsorption properties and hydration forces. However, since mica is not a simple protic surface characterised by a single pK value, the relation between solution conditions and the adsorption equilibria are ion-specific and quite complex\cite{DeYoreo2020}. Here, we report surface force measurements between mica sheets across aqueous solutions of \ce{NaHCO3} and \ce{Na2CO3} at different concentrations, and compare these to literature measurements for \ce{KCl}. We consider the fitted effective surface charge, screening length, and hydration forces observed. Each of these appears to be distinctly different in the bicarbonate-containing solutions compared to \ce{KCl}; we discuss the role of speciation on surface and colloidal interactions, and make connections to recent grand-canonical Monte-Carlo simulations of surface forces in a similar charge regulating system. \\

\section*{Methods}
The interaction force, $F$, as a function of separation distance, $D$, between atomically smooth mica sheets in crossed-cylinder configuration and immersed in aqueous solutions was measured using a Surface Force Balance (SFB)\cite{Hayler_2024}.
When the radius of curvature of the mica cylinders, $R$, is much greater than the separation between them, \textit{i.e.} $R>>D$, we can write that  $F/2\pi R = W^{||}$ where $W^{||}$ is the free energy of interaction (potential of mean force) between parallel sheets, per unit area, at separation $D$ (the Derjaguin approximation). This is essentially exact in our case where $R \approx 1$~cm, $D\approx 1-100$~nm.  For comparison to theory and simulations, $W^{||}$ is the excess grand potential since the electrolyte between the mica sheets is in (electro-)chemical and thermal equilibrium with the reservoir\cite{Evans1987}. \\
White light interferometry is used to determine $D$ (equivalent to the electrolyte film thickness) to 0.1 nm precision and 0.2-0.5 nm accuracy relative to mica-mica contact; in the present work we present measurements of distance relative to the closest approach within an experiment which we denote $D_0$. Forces are measured via deflection of a spring in the direction perpendicular to the crossed cylinders' axes. A schematic of the SFB is provided in Figure~\ref{fig:1}(b). Interactions were measured from $\approx 200$~nm down to contact, however we present only the region at shorter distances where non-zero forces were detected. The detailed design and operating principles of the apparatus have been described extensively elsewhere\cite{Hayler_2024}.
Details specific to the present study are as follows. Ruby muscovite mica sheets (S\&J Trading Inc.) were cleaved in a particle-free environment to ensure atomically smooth surfaces. The mica pieces used had thicknesses between 2-5 $\unit{\micro\metre}$ and were back-silvered with a 45 nm thick silver layer prior to being mounted on cylindrical glass lenses using EPON glue.
Electrolyte solutions consisted of sodium bicarbonate (\ce{NaHCO_3}) and sodium carbonate (\ce{Na_2CO_3}) salts obtained in anhydrous form from Fisher Scientific UK Ltd. with a purity of 99.99\% (Puratronic\textsuperscript{\textregistered}) and potassium chloride (\ce{KCl}) obtained from Thermo Scientific with a purity of 99.997\% (Puratronic\textsuperscript{\textregistered}). The water used was generated from a Milli-Q high-purity system, with the total organic carbon (TOC) < 4 ppb and the resistivity of 18.2 M${\unit{\ohm}}$cm. Solutions were prepared at concentrations of 1 mM for \ce{KCl}, 1 mM and 10 mM for \ce{NaHCO_3}, and 10 mM for \ce{Na_2CO_3}. The pH of each electrolyte solution was measured using a calibrated pH meter (HI5221, HANNA\textsuperscript{\textregistered} Instruments)\\

\section*{Results and Discussion}

\subsection*{General features of the force-distance profiles}
\begin{figure}[h!]
    \centering
    \includegraphics[width=0.5\linewidth]{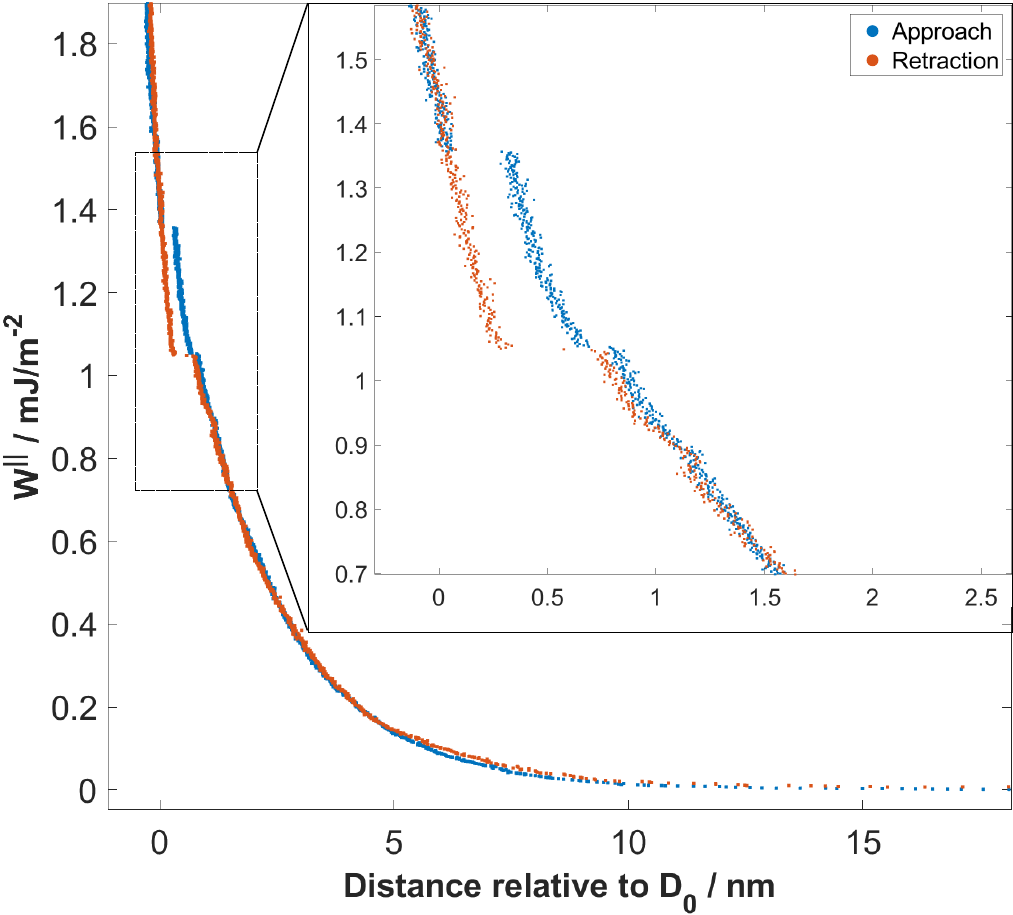}
    \caption{Interaction free energy per unit area, $W^{||} = F/2\pi R$, between mica sheets as a function of their separation distance (relative to closest separation), $D_0$, measured across 10 mM \ce{Na2CO3} aqueous solution at 294 K. Interactions measured on approach (blue curve) and retraction (red curve) are shown. The force profiles are reversible in the range down to 1~nm, then small steps and a hysteresis are seen (enlarged in the inset). At the smallest separations the force is again reversible, showing absence of adhesion.}
    \label{fig:2}
\end{figure}

The interaction force ($F$) as a function of separation distance ($D$), was measured between atomically smooth mica surfaces immersed in aqueous solutions of sodium bicarbonate and sodium carbonate for various bulk concentrations in the range 1-100 mM. Forces were measured on both approach and retraction of the surfaces, for distances in the range $\approx 200$~nm down to 0~nm. A representative force-distance profile is shown in Figure \ref{fig:2}, for 10 mM \ce{Na2CO3}. The mica surfaces experience a repulsive force, measurable from approximately 10-15~nm, which increases exponentially with decreasing separation down to $\approx$ 2~nm. Below 2~nm, small jumps inwards are observed, of size 0.1~nm then 0.3~nm.  On retraction of the surfaces a hysteresis is observed in the region of the jumps (spring instabilities) but the force profile becomes reversible again beyond 1-2~nm. At contact or near-contact separations, $0<D<0.2$~nm, the interaction force remains repulsive at all pressures applied; there is no adhesion. Qualitatively similar features were observed for all bicarbonate electrolytes studied, and can be interpreted in terms of (reversible) mean-field DLVO forces at longer range and structural forces, or \textit{hydration forces}, at short range arising from the discontinuous squeeze-out of hydrated ions from the film between the mica sheets. In the following we discuss quantitive differences observed between bicarbonate solutions at varying concentration and pH, and their comparison to simple salt solutions.    

\subsection*{Bicarbonate in the reservoir adjusts the effective surface charge}
\begin{figure}[h!]
    \centering
    \includegraphics[width=1\linewidth]{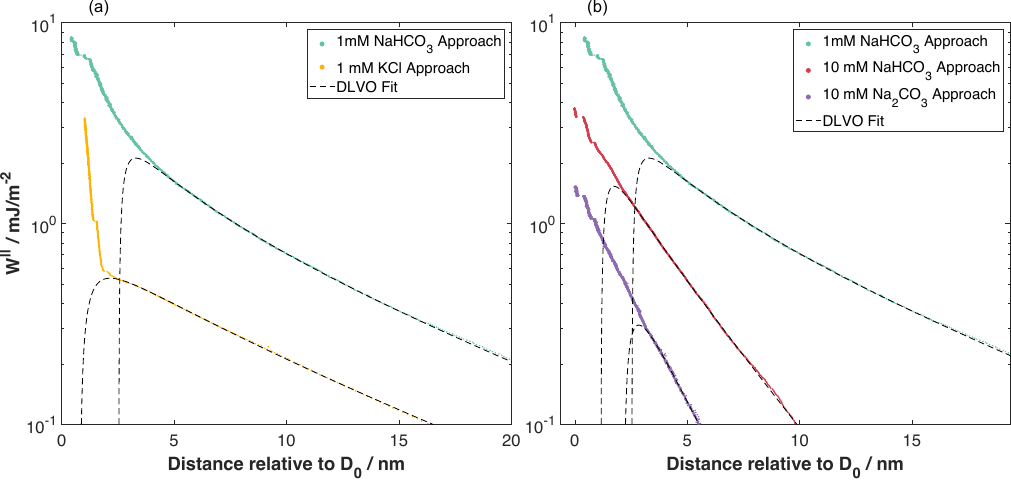}
    \caption{Interaction energy per unit area measured between mica sheets as a function of separation distance. (a) 1 mM \ce{NaHCO3} compared to 1 mM \ce{KCl}\cite{Agg2025}. (b) 1 mM \ce{NaHCO3} compared to 10 mM \ce{NaHCO3} and 10 mM \ce{Na2CO3}. In all cases the dashed lines are fits to equation~\ref{eqn:DLVO}, with fitting parameters in Table~\ref{tab:1}.}
    \label{fig:3}
\end{figure}

Figure \ref{fig:3}(a) compares the measured interaction force across 1 mM \ce{NaHCO3} to the interaction force across 1 mM \ce{KCl}. The range of the interaction, captured by the exponential decay length, is similar for the two electrolytes. However, the magnitude of the force is substantially larger in the case of \ce{NaHCO3} despite the same 1~mM bulk concentration. To interpret these observations we consider a simple model for the mean-field interaction force between symmetric charged surfaces across electrolyte, involving a sum of van der Waals and electrostatic forces. Electrostatic forces are modelled using the Poisson-Boltzmann equation, expected to be suitable at this low bulk ion concentration, with allowance for charge regulation as the surfaces approach. To account for the charge regulation we use a constant regulation approximation\cite{Trefalt2016,Smith2020}, involving a regulation parameter, $p$, taking values between 0 and 1 to interpolate between the limits of constant potential (CP; $p=0$) and constant charge (CC; $p=1$)\cite{Carnie1993,PericetCamara2004}. The total interaction force including the van der Waals term is:

\begin{equation}
	\frac{F}{2\pi R} =W^{||}= -\frac{A}{12{\pi}D^2} +  2\epsilon_0\epsilon_\textrm{r}{\kappa}\psi_{\text{eff}}^2\frac{e^{-{\kappa}D}}{1+(1-2p)e^{-{\kappa}D}}
	\label{eqn:DLVO}
\end{equation}
where $A$ is the Hamaker constant, $\epsilon_0$ is the permittivity of free space, $\epsilon_\textrm{r}$ is the relative permittivity, $\psi_{\text{eff}}$ is the mica effective surface potential (at large surface separation), $\kappa^{-1}$ is the screening length. $F$, $R$ and $D$ are measured in the experiment, $A$ and $\epsilon_0\epsilon_\textrm{r}$ are known for the system, and $\psi_{\text{eff}}$, $\kappa$ and $p$ are treated as fitting parameters. $\kappa^{-1}$ may also be compared to the theoretical Debye-H{\"{u}}ckel screening length, $\kappa_{\textrm{D}}^{-1}$ (see below). The fitted values for these force profiles are shown in Table~\ref{tab:1}. The fitted screening length $\kappa^{-1}$ is the same within error for the two electrolytes and is in accordance with the calculated Debye-H{\"{u}}ckel screening length for a 1:1 electrolyte at 1 mM concentration. However, there are two clear differences between the measured force profiles for 1 mM \ce{NaHCO3} and 1 mM \ce{KCl}: (i) 1 mM \ce{NaHCO3} has a larger fitted $\psi_{\text{eff}}$, and (ii) the fitted $p$-values imply almost constant charge boundary condition for 1 mM \ce{NaHCO3} in contrast to \ce{KCl} which facilitates charge regulation. We will consider these two observations in turn. \\

\begin{table}[h!]
    \centering
    \caption{Fitting parameters for the datasets shown in Figure~\ref{fig:3}. These fitting parameters and associated errors are relevant for the individual profiles; in the SI we provide fitting parameters for the multiple force-runs carried out in each solution including across multiple experiments (different mica sheets).}
    \label{tab:1}
    \begin{tabular*}{\textwidth}{@{\extracolsep{\fill}}lllll}
        \hline
        Parameter & 1 mM $\ce{NaHCO_3}$ & 10 mM $\ce{NaHCO_3}$ & 10 mM $\ce{Na_2CO_3}$ & 1 mM KCl \\ 
        \hline
        $\kappa^{-1}$ / nm & 10.2 $\pm$ 0.2 & 3.2 $\pm$ 0.05 & 2.2 $\pm$ 0.1 & 10.0 $\pm$ 0.2 \\
        $p$ & 0.85 $\pm$ 0.10 & 0.74 $\pm$ 0.04 & 0.77 $\pm$ 0.09 & 0.75 $\pm$ 0.04 \\
        $\psi_{\textrm{eff}}$ / mV & 87 $\pm$ 10 & 64 $\pm$ 3 & 35 $\pm$ 3 & 58 $\pm$ 2 \\ 
        \hline
    \end{tabular*}
\end{table}

The difference in $\psi_{\textrm{eff}}$ between 1 mM \ce{KCl} and 1 mM \ce{NaHCO3} is unlikely to arise from the $\ce{K+/Na+}$ cations\cite{Pashley1981b}. Instead, the origin of the striking difference in $\psi_{\textrm{eff}}$ is likely attributed to the higher reservoir pH of 1 mM \ce{NaHCO3}; the bicarbonate speciation equilibria, equation~\ref{eqn:bicarbonate}, act as a sink for protons and cations, driving the mica surface adsorption equilibria in the direction of more negative surface charge. To illustrate this, a simplistic model for the mass-action equilibria involving mica and its counterions is shown in equation~\ref{eqn:mica}. The equilibrium involves mica neutralised by adsorbed ions (here \ce{M+} implies metal, e.g. \ce{Na+} or \ce{K+}), mica with hydrated ions adsorbed (often called "Stern layer"), and mica holding a negative charge due to dissolution of counterions into the diffuse part of the electrical double layer. The mica equilibrium is distinct from that of a protic surface such as silica, and mica is not normally considered to have a simple isoelectric point or surface acidity constant. Nonetheless adsorption and surface charge respond to the bulk electrolyte conditions (concentration, pH) through multiple coupled processes.  \\

We are not aware of calculations or simulations of the relevant mica equilibria in the presence of bicarbonate or other pH-buffering solution, although suitable methods have recently been devised and applied to similar scenarios\cite{Levin2023,Levin2025} as discussed below. Complicating factors include the variation in \ce{H+} concentration in the electrical double layer vs. bulk solution due to the requirement for constant electrochemical potential between the reservoir and confined/surface regions\cite{Levin2023Lang} and solvent dielectric variations between the Stern layer and bulk. \\

Nonetheless, it is clear that the mica surface equilibria are coupled to the bicarbonate bulk speciation equilibria, equation~\ref{eqn:bicarbonate}, through the hydronium ion: an increase in pH due to the presence of bicarbonate in the reservoir (away from the surfaces) drives the mica equilibrium in the direction of more negative charge and higher negative surface potential. 

\begin{equation}\label{eqn:mica}
\begin{split}
\mathrm{mica}-\ce{M} +n \mathrm{H}_2\mathrm{O} \mathrm{(l)} &\rightleftharpoons \mathrm{mica}^--\ce{M(H2O)}_n^+ 
\rightleftharpoons \mathrm{mica}^- + \ce{M(H2O)}_n^+ \mathrm{(aq)} \\
\mathrm{mica}-\ce{H} +\mathrm{H}_2\mathrm{O} \mathrm{(l)} &\rightleftharpoons
\mathrm{mica}^--\ce{H3O+}  \rightleftharpoons \mathrm{mica}^- + \ce{H3O+} \mathrm{(aq)}
    \end{split}
\end{equation}

While the mica-charging and $\psi_{\textrm{eff}}$ effects discussed above apply to mica-solution interfaces in general (no confinement; $D=\infty$), charge regulation is intrinsically a confinement effect ($D\rightarrow0$). The fitted $p$-values for 1 mM \ce{NaHCO3} and 1 mM \ce{KCl} reveal differences in charge regulation, with 1 mM \ce{NaHCO3} close to CC and  1 mM \ce{KCl} displaying significant CR. This is directly apparent on inspection of the force profiles: the curvature of $\ln(W)$ vs. $D$ below about 10~nm (\textit{i.e.} $D< \kappa^{-1}$) across 1 mM \ce{NaHCO3} is convex, deviating from the plain exponential decay at larger distances where the boundary condition has little effect on the functional form. As $D$ decreases in the range $D< \kappa^{-1}$, the boundary condition (CC or CR) has an increasingly strong influence on the interaction. The molecular origin of CR, for 1~mM \ce{KCl}, involves adsorption of diffuse-layer ions to the surfaces as $D$ decreases to lower the overall free energy of interaction. In 1 mM \ce{NaHCO3}, on the other hand, there is little or no charge regulation implying that ions are not driven to adsorb on the surfaces during compression of the film; the surface charge remains constant (and high), giving rise to a steeper increase in excess pressure. Looking again at the simple mass-action expressions in equation~\ref{eqn:mica}, we can interpret that, for \ce{KCl}, the position of equilibrium shifts to the left as two mica surfaces approach whereas in \ce{NaHCO3} this is not the case. It may be that, for \ce{KCl} solutions, adsorption of \ce{H3O+} onto the mica or incorporation of \ce{H+} within it provide the mechanism for CR whilst this mechanism is not available at high pH and low ionic strength. A supporting counterexample is the observation that acidic solutions displayed even stronger reduction of the mica surface charge during approach of two surfaces leading to attraction and adhesion\cite{Pashley1981a}. \\

The effect of bulk sodium bicarbonate concentration is illustrated by comparing the 1 mM \ce{NaHCO3} measurement to 10 mM \ce{NaHCO3} in Figure~\ref{fig:3}(b) and corresponding fitting parameters in Table~\ref{tab:1}. With increased \ce{NaHCO3} concentration the surface potential is lower, the range (screening length) is shorter, and the boundary condition best fitting the measurement is with substantial charge regulation. It appears that, with a higher concentration of \ce{Na+} and \ce{HCO3-} in solution, charge regulation is now favoured during surface approach; the 10-fold increase in solution \ce{Na+} concentration appears sufficient to shift the equilibrium fraction of charged mica sites. Although $\psi_{\textrm{eff}}$ is lower in 10 mM compared to 1 mM \ce{NaHCO3}, it remains substantially higher than in \ce{KCl} or other 1:1 salt solutions at similar concentration. Therefore the influence of high pH on effective charge and $\psi_{\textrm{eff}}$ remain at this higher concentration. The decrease in screening length is anticipated within the Debye-H{\"{u}}ckel theory; quantitative effects of the speciation involved are discussed in the next section. \\

\subsection*{Screening length depends on carbonate speciation}
In Figure~\ref{fig:3}(b) we make a further comparison between the mica-mica interaction free energy across solutions of sodium bicarbonate (\ce{NaHCO3}) and sodium carbonate (\ce{Na2CO3}). As anticipated by the speciation equilibria for bicarbonate solutions in equation~\ref{eqn:bicarbonate}, addition of sodium bicarbonate tips the equilibria to the right and the resulting solution has higher pH; the measured pH and absolute concentrations of each species in the equilibrium are provided for our experimental scenarios in Table~\ref{tab:2}.  The impact of this shift in the speciation equilibria on surface forces is clear: we see that the absolute magnitude of the repulsion is lower, and the screening length $\kappa^{-1}$ is also shorter, for \ce{Na2CO3} compared to \ce{NaHCO3} at the same 10 mM reservoir concentration. To rationalise the lower $\kappa^{-1}$ in \ce{Na2CO3}, we calculate the theoretical Debye screening length:  

\begin{equation}
\kappa_\textrm{D}^{-1}=\bigg(\frac{\epsilon_\textrm{r} \epsilon_0 k_\textrm{B}T}{\frac{1}{2}\sum_i c_ie^2z_i^2}\bigg)^{1/2}
	\label{eqn:Debye}
\end{equation}
with $e$ the electron charge, $c_i$ the concentration of ionic species $i$ and $z_i$ its valence. Taking the sum over all ionic species present: \ce{Na+}, \ce{HCO3-}, \ce{CO3^{2-}} (with \ce{H+} and \ce{OH-} having negligible concentration), and calculating their concentrations using the equilibrium constants for speciation as in equation~\ref{eqn:bicarbonate}, we calculate theoretical screening lengths as shown in Table~\ref{tab:2}. The total ionic strength, $I=\frac{1}{2}\sum c_iz_i^2$, is about three times higher in \ce{Na2CO3} due to almost 100 times higher concentration of \ce{CO3^{2-}} and the strong influence of divalent ions on effective ionic strength of an electrolyte. The result of this is a small decrease in the predicted (theoretical) screening length, $\kappa_\textrm{D}^{-1}$ from 3.0~nm to 1.7~nm, which is closely similar to the decrease seen in the experimental measurements in Figure~\ref{fig:3}(b).  This effect of speciation on the observed screening length, well anticipated in classic electrolyte theory, was clear and reproducible for the single example electrolyte and concentrations reported here. However we have not yet extended our studies to higher concentrations, where non-ideality (not taken into account in equation~\ref{eqn:Debye} or our calculation) is likely to be more significant, or to other buffering species.    

\begin{table}[h!]
    \centering
    \caption{Theoretical screening length, $\kappa_\textrm{D}^{-1}$, and speciation corresponding to the measured pH of the solutions measured and shown in Figure \ref{fig:3}.}
    \label{tab:2}
    \resizebox{\textwidth}{!}{%
    \begin{tabular}{lllllllllllll}
        \hline
        Solution & pH & $f_{\ce{H_2CO_3}} (\%)$ & $f_{\ce{HCO_3^-}} (\%)$ & $f_{\ce{CO_3^{2-}}} (\%)$ & Total DIC (M) & $c_{\ce{H_2CO_3}}$ (M) & $c_{\ce{HCO_3^-}}$ (M) & $c_{\ce{CO_3^{2-}}}$ (M) & Ionic strength (M) & $\kappa_\textrm{D}^{-1}$ (nm) \\ 
        \hline
        1 mM NaHCO$_3$ & 8.03 & 2.11 & 97.39 & 0.5 & 1 $\times 10^{-3}$ & 2.11 $\times 10^{-5}$ & 9.74 $\times 10^{-4}$ & 5.02 $\times 10^{-6}$ & 9.98 $\times 10^{-4}$ & 9.6 \\
        10 mM NaHCO$_3$ & 8.20 & 0.00 & 97.82 & 0.75 & 1 $\times 10^{-2}$ & 1.43 $\times 10^{-4}$ & 9.78 $\times 10^{-3}$ & 7.46 $\times 10^{-5}$ & 1.00 $\times 10^{-2}$ & 3.0 \\ 
        10 mM Na$_2$CO$_3$  & 11.6 & 0.00 & 4.96 & 95.04 & 1 $\times 10^{-2}$ & 2.89 $\times 10^{-9}$ & 4.96 $\times 10^{-4}$ & 9.50 $\times 10^{-3}$ & 3.12 $\times 10^{-2}$ & 1.7 \\ 
        \hline
    \end{tabular}
    }
\end{table}

\subsection*{Short-range hydration forces in carbonate/bicarbonate systems}

\begin{figure}[h!]
    \centering
    \includegraphics[width=1\linewidth]{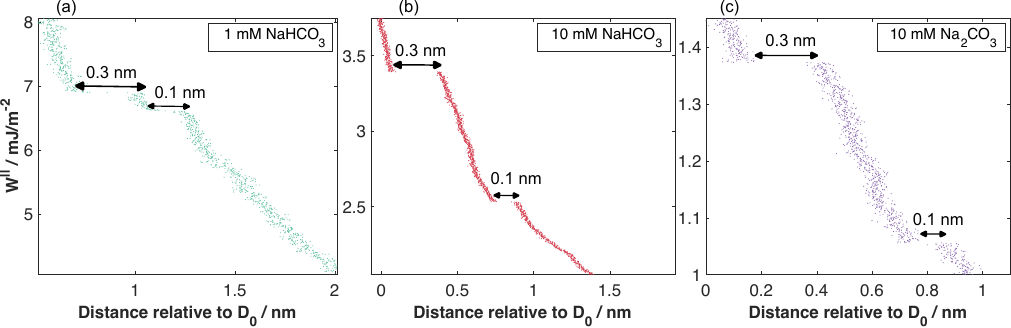}
    \caption{High resolution measurement of the interaction as a function of separation below $\approx2$~nm, in (a)-(c) for each of the solutions discussed above, showing discontinuities attributed to structural changes in the film. In each case a step of 0.1~nm precedes a step of 0.3~nm.}
    \label{fig:4}
\end{figure}

At separations below $\approx$ 2-3~nm, the force-distance profiles deviate substantially from the mean-field behaviour described in equation~\ref{eqn:DLVO} and instead displayed discrete, reproducible steps in the measured separation indicative of structural layering transitions in the confined liquid. Two characteristic step sizes were identified: one of approximately 0.1~nm and another of 0.3~nm. These are shown in Figure~\ref{fig:4}, with one example from each of the bicarbonate solutions studied.  The smaller 0.1~nm step consistently appeared at slightly larger separations, preceding the 0.3~nm transition during approach.
The 0.3~nm step corresponds closely to the molecular diameter of a water molecule and can be attributed to the expulsion of a single hydration layer between the approaching mica surfaces, a phenomenon commonly observed in SFB and AFM studies of aqueous systems and atributed to structural hydration forces. The smaller 0.1~nm step, however, is unusual. Its reproducibility across multiple measurements and solution conditions, always when bicarbonate-containing electrolytes are present, suggests a genuine interfacial event rather than a mechanical or instrumental artefact. The origin of this 0.1~nm step remains uncertain, one possible explanation is the partial displacement or rearrangement of hydrated protons or tightly bound ion-water complexes near the charged mica surface. The fact that this step always precedes the water-layer squeeze-out supports the idea that subtle protonic or hydration-shell reorganizations occur before complete dehydration of the confined region. Further experimental and theoretical work will be required to uncover its origin.

\section*{Conclusions}

We have presented initial results for the interaction free energy, derived from direct force measurements, between muscovite mica sheets across aqueous solutions of carbonate and bicarbonate salts and in equilibrium with bulk solution. The results are explained using classical theories for colloidal interactions incorporating electrostatic/entropic and van der Waals terms, except for the hydration forces observed at very short distances. Quantitative comparison to simple salt solutions reveals a strong influence of pH on the surface effective potential and, at low concentrations, the absence of charge-regulation during approach of the surfaces. At 1-2 nm range, the hydration forces also displayed different features.  \\
Recently, grand-canonical Monte-Carlo (GCMC) simulation methods have been developed to handle electrolytes confined between charged surfaces comprised of individual ionisable groups and in equilibrium with a bulk reservoir\cite{Levin2023,Levin2023Lang,Levin2025}. There it was shown that the force between the surfaces increases more rapidly with decreasing distance for solutions at higher pH, and furthermore that surfaces held at constant charge boundary condition led to steeper repulsion than charge regulating surfaces\cite{Levin2025}. The GCMC findings are in accordance with our initial measurements presented here for bicarbonate solutions between mica surfaces, although we note that the simulated surfaces involved a simpler proton-exchange equilibrium compared to the complex system of physisorption normal for aprotic mineral surfaces such as mica. The expeirments presented here represent an initial data set and further measurements are required, in partiuclar to approach higher concentrations of both bicarbonate and background salt.\\ 
Our initial findings highlight how speciation of bicarbonates in bulk solution influences interfacial forces and molecular organization in confined electrolytes. Even though the negatively charge carbonate and bicarbonate ions are themselves present at only negligible concentration in the confined region due repulsion from the negatively charged bounding surfaces, they act strongly to influence the pressure and structural forces via the electrochemical potential equilibria between reservoir and film. 
These findings, and future experiments exploring the bicarbonate system in more detail, may be relevant towards understanding carbonate speciation effects on ion transport and stability in natural and technological aqueous environments\cite{CrystGrow_Meldrum2025}.

\section*{Data availability}
All raw data is provided at the following DOI [to be provided], supported by the Oxford Research Archive. 

%%%FOOTNOTES%%%

\footnotetext{Physical and Theoretical Chemistry Laboratory, Department of Chemistry, University of Oxford, Oxford OX1 3QZ, UK. E-mail: susan.perkin@chem.ox.ac.uk}

%Please use \dag to cite the ESI in the main text of the article.
%If you article does not have ESI please remove the the \dag symbol from the title and the footnotetext below.
\footnotetext{\dag~Supplementary Information available: additional measurement runs of force profiles and derivation of species fractions for the carbonate system. See DOI: 00.0000/00000000.}
%additional addresses can be cited as above using the lower-case letters, c, d, e... If all authors are from the same address, no letter is required

%\footnotetext{\dag~Additional footnotes to the title and authors can be included \textit{e.g.}\ `Present address:' or `These authors contributed equally to this work' as above using the symbols: \ddag, \textsection, and \P. Please place the appropriate symbol next to the author's name and include a \texttt{\textbackslash footnotetext} entry in the the correct place in the list.}

%%%END OF FOOTNOTES%%%

\section*{Author Contributions}
S.P., S.M. and D.T. conceived and planned the experiments. D.T., S.M., and K.A. performed the experiments and fitting. S.P. D.T., S.M., K.J.A. interpreted the data and wrote the paper. 

\section*{Conflicts of interest}
There are no conflicts to declare.

\section*{Acknowledgments}
 The authors gratefully acknowledge funding from the European Research Council under grant 101001346 ELECTROLYTE. S.M. is supported by a Career Development Research Fellowship from St. John's College, Oxford. K.J.A. would like to acknowledge support from The Oxford-The Queen’s College Graduate Scholarship in partnership with the Clarendon Fund, University of Oxford.

\bibliography{rsc} 
... \\\\

\bibliographystyle{rsc} %the RSC's .bst file

\end{document}